\documentclass[preprint,superscriptaddress,amsmath,amssymb,prd,aps,showpacs,floatfix,nofootinbib]{revtex4-1}
\usepackage{graphicx}
\usepackage{dcolumn}
\usepackage{bm}
\usepackage{mathrsfs}
\usepackage{hyperref}
\usepackage{amsmath}
\usepackage{amssymb}
\usepackage{amsfonts}
\usepackage{xcolor}
\usepackage{float}
\usepackage{dcolumn}
\usepackage{multirow}

\begin{document}
\title{The $X(3960)$ seen in $D_{s}^{+} D_{s}^{-}$ as the  $X(3930)$ state seen in $ D^{+} D^{-} $ }

\date{\today}
\author{M.~Bayar}
\email{melahat.bayar@kocaeli.edu.tr}
\affiliation{Department of Physics, Kocaeli University, Izmit 41380, T{\"u}rkiye}
\affiliation{Departamento de
F\'{\i}sica Te\'orica and IFIC, Centro Mixto Universidad de
Valencia-CSIC Institutos de Investigaci\'on de Paterna, Aptdo.22085,
46071 Valencia, Spain}

\author{A.~Feijoo}
\email{edfeijoo@ific.uv.es}
\affiliation{Departamento de
F\'{\i}sica Te\'orica and IFIC, Centro Mixto Universidad de
Valencia-CSIC Institutos de Investigaci\'on de Paterna, Aptdo.22085,
46071 Valencia, Spain}

\author{E.~Oset}
\email{oset@ific.uv.es}
\affiliation{Departamento de
F\'{\i}sica Te\'orica and IFIC, Centro Mixto Universidad de
Valencia-CSIC Institutos de Investigaci\'on de Paterna, Aptdo.22085,
46071 Valencia, Spain}

\begin{abstract}

We perform a calculation of the interaction of the $ D \bar{D} $, $ D_{s} \bar{D}_{s} $ coupled channels and find two bound states, one coupling to $ D \bar{D} $ and another one at higher energies coupling mostly to $D_{s}^{+} D_{s}^{-}$. We identify this latter state with the  $X_{0}(3930)$ seen in the  $D^{+} D^{-}$  mass distribution in the $B^+ \to D^{+} D^{-} K^{+} $ decay, and also show that it produces an enhancement of the  $D_{s}^{+} D_{s}^{-}$ mass distribution close to threshold which is compatible with the LHCb recent observation in the  $B^+ \to D_{s}^{+} D_{s}^{-} K^{+} $ decay which has been identified as a new state, $X_{0}(3960)$.

\end{abstract}

\maketitle


\section{Introduction}
In a recent talk at CERN, the LHCb Collaboration reported on three new states \cite{cern} one of which, named as $X(3960)$, is seen as a peak in the $D_{s}^{+} D_{s}^{-}$ mass distribution of the $B^+ \to D_{s}^{+} D_{s}^{-} K^{+} $ decay.  The properties assigned to that state are 

\begin{align}
J^{PC}=0^{++}~; ~~M_{0}=3955 \pm 6 \pm 11 ~MeV ~ ; ~~\Gamma_{0}=48 \pm 17 \pm 10 ~MeV  \nonumber
\end{align}

The peak is remarkably close to the $D_{s}^{+} D_{s}^{-}$  threshold, $ 3937 $ MeV, which makes one wonder where it could not be a signal for a resonance just below threshold \footnote{Comment of Sasa Prelovsek in the discussion of the talk  \cite{cern}}. This is actually a well known feature of reactions as discussed in \cite{guozou} and found in some specific reactions \cite{navarra,enwang}. Actually, the $B^+$ decay into $B^+ \to D^{+} D^{-} K^{+} $ showed a signal in the  $D^{-} D^{+}$ mass distribution for a $ 0^{++} $ state, branded $X_{0}(3930)$, with properties \cite{dpdm1,dpdm2}

\begin{align}
J^{PC}=0^{++}~; ~~M'_{0}=3924 \pm 2 ~MeV ~~ ;~~  \Gamma'_{0}=17 \pm 5 ~MeV  \nonumber
\end{align}

If the $X_{0}(3930)$ state coupled both to  $D^{+} D^{-}$ and $D_{s}^{+} D_{s}^{-}$, that state would necessarily produce an enhancement close to threshold in the $D_{s}^{+} D_{s}^{-}$ mass distribution, which could explain the experimental observation without the need to introduce an extra resonance. The purpose of this work is to show that present dynamics of the interaction of charmed mesons leads naturally to this conclusion. 

The first consideration in this respect is the QCD lattice result of \cite{sasa}, where a $ 0^{++} $ bound state coupling strongly to $D_{s}^{+} D_{s}^{-}$ and weakly to $D^{+} D^{-}$ is found below the $D_{s}^{+} D_{s}^{-}$ threshold. Such a state would indeed show a peak in the $D^{+} D^{-}$ mass distribution and an enhanced mass distribution of the   $D_{s}^{+} D_{s}^{-}$ around threshold. 

One might expect that such a state would appear in a dynamical study of $ D \bar{D} $ and $D_{s}^{+} D_{s}^{-}$ in coupled channels. Interestingly, this study has been done in \cite{dani, juandd, hidalgo}, where a $ D \bar{D} $ bound state was found. Yet, no bound state was found close to the $D_{s}^{+} D_{s}^{-}$ threshold coupling mostly to that channel.

In what follows we show that this is a consequence of the strong $ D \bar{D} \to D_{s}^{+} D_{s}^{-}$ transition, to the point that a little weaker transition already gives rise to the two states, the upper one coupled mostly to $D_{s}^{+} D_{s}^{-}$ as in \cite{sasa} and with properties that are consistent with the mass and width of the  $X_{0}(3930)$ and the strength and shape of the $D_{s}^{+} D_{s}^{-}$ mass distribution of the $X_{0}(3960)$. A formulation of the problem is presented below. 

\begin{figure}[h!]
  \centering
  \includegraphics[width=0.50\textwidth]{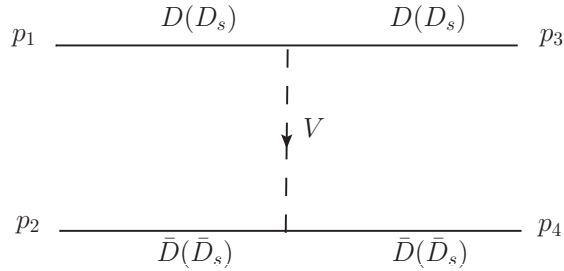}
  \caption {Dynamics of $ D (D_{s}) \to \bar{D} (\bar{D}_{s})$ interaction due to vector exchange. }
   \label{feyn_DDS}
\end{figure}

We depart from the dynamics of the  $ D \bar{D} $, and coupled channels used in \cite{dani} and use a formulation based on the extension of the local hidden gauge approach \cite{hidden1,hidden2,hidden4,hideko}, which has turned out to make accurate predictions for $D^{(*)} K^{(*)}$, $D^{(*)} \bar{K}^{(*)}$ states \cite{branz, raquel} or $ T_{cc} $ and related states \cite{feijoo,dai} among others. The dynamics is based on the exchange of vector mesons as depicted in Fig. \ref{feyn_DDS}

One needs the $ VPP $ ($ V= $ vector, $ P= $pseudosclar) vertex given by 

\begin{equation}
{\cal L}_{VPP}=-ig\langle [P,\partial_\nu P]V^\mu \rangle; ~g=\dfrac{M_V}{2 f} ~(M_V\simeq800 ~MeV, f=93 ~MeV) \label{eq:LAG}
\end{equation}
where $ P $ and $ V $ are the $ q_{i} \bar{q}_{j}$ matrices written in terms of pseudoscalar  ($ P $)  and vector  ($ V $) mesons respectively

\begin{equation}\label{Psematrix}
P = \begin{pmatrix}
\frac{1}{\sqrt{2}}\pi^0 + \frac{1}{\sqrt{3}} \eta + \frac{1}{\sqrt{6}}\eta' & \pi^+ & K^+ & \bar{D}^0 \\
 \pi^- & -\frac{1}{\sqrt{2}}\pi^0 + \frac{1}{\sqrt{3}} \eta + \frac{1}{\sqrt{6}}\eta' & K^0 & D^- \\
 K^- & \bar{K}^0 & -\frac{1}{\sqrt{3}} \eta + \sqrt{\frac{2}{3}}\eta' & D_s^- \\
D^0  & D^+ & D_s^+ & \eta_c
\end{pmatrix},
\end{equation}

\begin{equation}\label{Vecmatrix}
V = \begin{pmatrix}
 \frac{1}{\sqrt{2}}\rho^0 + \frac{1}{\sqrt{2}} \omega & \rho^+ & K^{* +} & \bar{D}^{* 0} \\
 \rho^- & -\frac{1}{\sqrt{2}}\rho^0 + \frac{1}{\sqrt{2}} \omega  & K^{* 0} & \bar{D}^{* -} \\
 K^{* -} & \bar{K}^{* 0}  & \phi & D_s^{* -} \\
 D^{* 0} & D^{* +} & D_s^{* +} & J/\psi
\end{pmatrix},
\end{equation}
and $\langle  ~~  \rangle  $ means the trace of the $ q_{i} \bar{q}_{j}$ matrices. A straightforward calculation of the diagrams of Fig. \ref{feyn_DDS} with the Lagrangian  of Eq. (\ref{eq:LAG}) considering the channels $ D \bar{D} $, $ I=0 $ and $D_{s}^{+} D_{s}^{-}$ with the $ (D^{+}, -D^{0}) ~(\bar{D}^{0}, D^{-})$ multiplets

\begin{equation}
(D \bar{D}, ~I=0 )=\frac{1}{\sqrt{2}} (D^{+}D^{-}+D^{0}\bar{D}^{0});~ D_{s}^{+} D_{s}^{-} \nonumber 
\end{equation}

gives the interaction potential 

\begin{equation}
V_{ij}= -B_{ij} g^{2} (p_{1}+p_{3})(p_{2}+p_{4})\label{eq:Vij}
\end{equation}
with

\begin{equation}\label{Bmatrix}
B= \begin{pmatrix}
 \dfrac{1}{2} \left( \dfrac{3}{M^{2}_{\rho}} +\dfrac{1}{M^{2}_{\omega}}+\dfrac{2}{M^{2}_{J/\Psi}} \right) & \sqrt{2} \dfrac{1}{M^{2}_{K^{*}}} \\
 \sqrt{2} \dfrac{1}{M^{2}_{K^{*}}} & \left( \dfrac{1}{M^{2}_{\phi}} +\dfrac{1}{M^{2}_{J/\Psi}} \right)  
\end{pmatrix}.
\end{equation}
where, projected over $S$-wave, we have 

\begin{equation}
(p_{1}+p_{3})(p_{2}+p_{4}) \to \dfrac{1}{2} \left[ 3 s-(m^{2}_{1}+m^{2}_{2}+m^{2}_{3}+m^{2}_{4})- \dfrac{1}{s} (m^{2}_{1}-m^{2}_{2})(m^{2}_{3}-m^{2}_{4})\right] 
\end{equation}

The unitarization via the Bethe-Salpeter equation gives the scattering matrix

\begin{equation}
T=  \left[ 1-VG\right]^{-1} V \label{eq:BetheSal}
\end{equation}
where G is the loop function for two meson intermediate states that we choose to regularize with dimensional regularization as we did in  \cite{dani}. 

An interesting thing happens with Eq. (\ref{eq:BetheSal}). If we eliminate the $ V_{12} $ term of Eq. (\ref{eq:Vij}) which connects the $D^{+} D^{-}$ and  $D_{s}^{+} D_{s}^{-}$ channels, we get two poles with ordinary values of the subtraction constant $ a\sim -1.5 $ of the $ G $ function in a wide range of values, meaning that the two  $ V_{11} $, $ V_{22} $  potentials are strong enough to bind the $ D \bar{D} $ and $ D_{s} \bar{D}_{s} $ components. Yet, when $ V_{12} $  is switched on, the $ D \bar{D} $ state remains and the $ D_{s} \bar{D}_{s} $ state disappears. This is what happened in \cite{dani} where the $ D_{s} \bar{D}_{s} $ state was not found. Curiously, if we weaken $ V_{12} $ to a value of about $ 70 \%$ of its strength in Eq. (\ref{Bmatrix}) the two states already appear.  Uncertainties of this type in our approach, together with the lattice results \cite{sasa} generating this state, prompt us to accept that level of uncertainty in order to obtain the two states and see if our hypothesis of  $X_{0}(3930)$ being responsible for the $X_{0}(3960)$ peak holds or not. 

\begin{figure}[h!]
  \centering
  \includegraphics[width=0.50\textwidth]{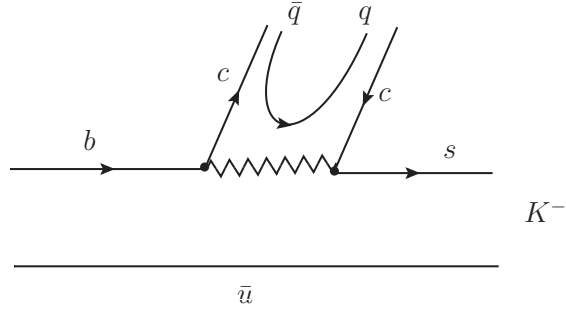}
  \caption{$ B^- $ decay via internal emission at the quark level and hadronization } 
   \label{feyn_BKcc}
\end{figure}

The next, non trivial, step is to relate the  $D^{+} D^{-}$ and $D_{s}^{+} D_{s}^{-}$ mass distributions. We use the charge conjugate reactions $B^- \to D^{+} D^{-} K^{-},~ D_{s}^{+} D_{s}^{-} K^{-}$  and have 

\begin{equation}
\dfrac{d~\Gamma}{dM_{\rm inv} }=\frac{1}{(2 \pi)^3} \frac{1}{4 M^{2}_{B}} ~ p_{K^-}~ p_{\tilde{D}_i}~\vert \tilde{t}_{i} \vert^{2}  \label{eq:dGDMinvDD} ,
\end{equation}
where $ p_{K^-} $ is the $ K^- $ momentum in the $ B^- $ rest frame, and $ p_{\tilde{D}_i} $ the $ D^{+} $ or $ D_{s}^{+} $ momenta in the $D^{+} D^{-}$, $D_{s}^{+} D_{s}^{-}$ rest frame respectively. The matrices $ \tilde{t}_{i} $ stand for the transitions $B^- \to D^{+} D^{-} K^{-} $ and $ D_{s}^{+} D_{s}^{-} K^{-} $ and are constructed in the following way. The  $ B^- $ decays proceed via interval emission \cite{chau}, as depicted in Fig. \ref{feyn_BKcc}, and are related. 

\begin{figure}[h!]
  \centering
  \includegraphics[width=0.80\textwidth]{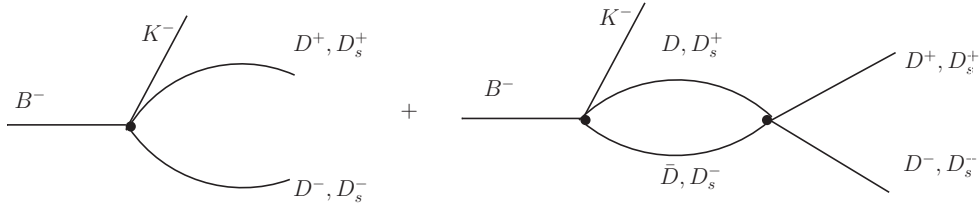}
  \caption{Production and propagation of the $D^{+} D^{-}$ and $D_{s}^{+} D_{s}^{-}$ components through final state interaction.}
    \label{feyn_BKDsDs}
\end{figure}

The $ c \bar{c} $ pair is hadronized and we have 

\begin{equation}
c \bar{c} \to \sum_{i} c \bar{q}_{i} q_{i}\bar{c} \to \sum_{i} P_{4i} P_{i4} =D^{0}\bar{D}^{0}+D^{+}D^{-}+D_{s}^{+} D_{s}^{-} = \sqrt{2} D \bar{D}+D_{s}^{+} D_{s}^{-}
 \label{eq:hadronization} ,
\end{equation}
where we have eliminated $ \eta_{c} \eta_{c}$ which plays no role here. Once $  D \bar{D} $ and $ D_{s}^{+} D_{s}^{-} $ have been created, they propagate as shown in Fig. \ref{feyn_BKDsDs}.

Then we immediately obtain 

\begin{equation}
\tilde{t}_{D^{+}D^{-}}=C\left(  1+ G_{D \bar{D}}(M_{inv}) T_{D \bar{D},D \bar{D}}(M_{inv}) +\dfrac{1}{\sqrt{2}} G_{D_{s} \bar{D}_{s}}(M_{inv}) T_{D_{s} \bar{D}_{s},D \bar{D}}(M_{inv}) \right) 
 \label{eq:DpDm} 
\end{equation}

\begin{equation}
\tilde{t}_{D_{s}^{+}D_{s}^{-}}=C \left(  1+ \sqrt{2} G_{D \bar{D}}(M_{inv}) T_{D \bar{D},D_{s}^{+} D_{s}^{-}}(M_{inv}) + G_{D_{s} \bar{D}_{s}}(M_{inv}) T_{D_{s}^{+} D_{s}^{-},D_{s}^{+} D_{s}^{-}}(M_{inv}) \right) 
 \label{eq:DspDsm} 
\end{equation}
with $ C $ an arbitrary constant. We can see that the $ D^{+}D^{-} $ and $D_{s}^{+} D_{s}^{-}$ production rates are related via the dynamics of the process and we can also test the relative strength of the two distributions.

We then proceed as follows. We choose a value of the subtraction constant $ a_{D_{s}\bar{D}_{s}} $ of $ G_{D_{s}\bar{D}_{s}} $ and a factor $ \alpha $ multiplying $ V_{12} $ such as to get approximately the right mass and width of the $X_{0}(3930)$ state.  We accomplish this with the values
 
 \begin{equation}
a_{D_{s} \bar{D}_{s}} =-1.58   ~; \alpha=0.7
 \label{eq:ads} 
\end{equation}
At the same time we choose $ a_{D \bar{D}} =-1 $ to get a $ D \bar{D} $ bound state around $ 3700 ~MeV $, as in \cite{dani}  \footnote{Values of $ a_{D \bar{D}} $ of the order of $ -1.3 $ were used in \cite{dani}  to get that binding, but also many other light channel were considered (see also \cite{xiao}) and it is well known that there is a certain trade off between term in the potential and changes in the $ G$ function.}, and we get the results in Table \ref{Tab:aDsDs}. These results  indicate that the lower state $ (I) $ couples more strongly to  $ D \bar{D} $  while the second state $ (II) $ couples more strongly to $ D_{s}\bar{D}_{s} $ as found in \cite{sasa}.

\begin{table*}

\centering
\caption{Masses and widths of the poles dynamically generated by the model, as well as, the corresponding modulus of the couplings  $|g_{i}|$. }
\label{Tab:aDsDs}
\begin{tabular*}{1.00\textwidth}{@{\extracolsep{\fill}} c c c c c}
\hline
\hline

          &  $M$ [MeV] & $\Gamma$ [MeV] & $|g_{\bar{D}D}|$   [MeV]  &  $|g_{\bar{D}_sD_s}| $ [MeV]    \\
 Pole I &  $3699$ &   $-$  &   $14516$     &  $5897$   \\
 Pole II (\textbf{$X_{0}(3930)$}) &  $3936$ &   $11$  &   $2858$     &  $9076$   \\

\hline
\hline

\end{tabular*}

\end{table*}

\begin{figure}
 \centering
 \includegraphics[scale=0.5]{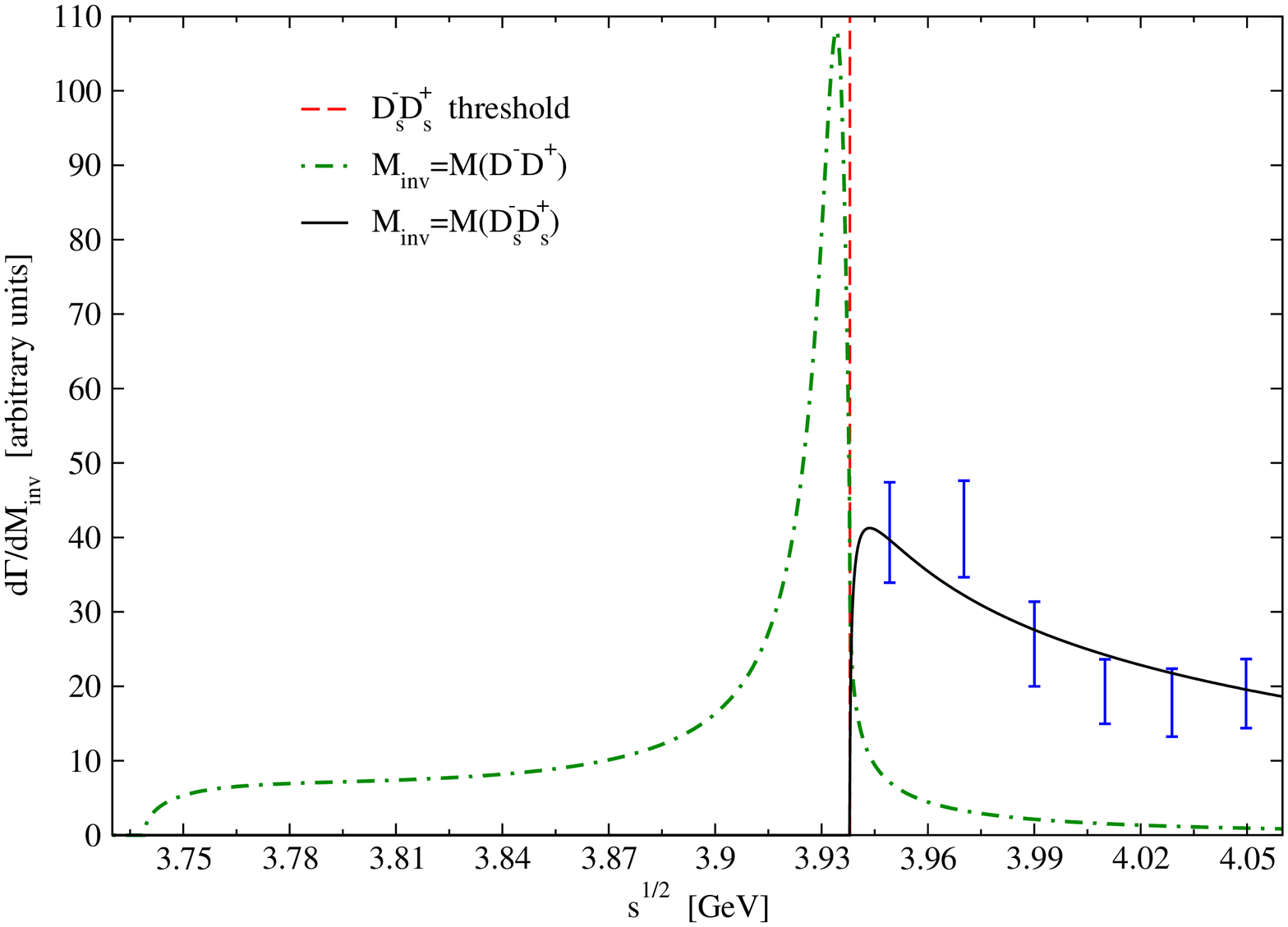}
 \caption{$ \dfrac{d~\Gamma}{dM_{\rm inv} (D^{+} D^{-}) } $ and  $ \dfrac{d~\Gamma}{dM_{\rm inv} (D_{s}^{+} D_{s}^{-}) } $ of  
  $B^- \to D^{+} D^{-} K^{-} $ and  $B^- \to D_{s}^{+} D_{s}^{-} K^{-} $  decays. The experimental points are taken from \cite{cern}.}
 \label{Result}
\end{figure}

In Fig. \ref{Result}, we show the results for $ \dfrac{d~\Gamma}{dM_{\rm inv} (D^{+} D^{-}) } $ and  $ \dfrac{d~\Gamma}{dM_{\rm inv} (D_{s}^{+} D_{s}^{-}) } $. The constant $ C $  has been chosen such as to have the normalization of the data for  $ \dfrac{d~\Gamma}{dM_{\rm inv} (D_{s}^{+} D_{s}^{-}) } $. We do not add a background to the distributions as in \cite{cern}, since our amplitude $\tilde{t}_{i}$ of Eqs. (\ref{eq:DpDm}), (\ref{eq:DspDsm}) already contain one through  the terms $ 1 $ in the parenthesis (tree level). We observe that with the chosen parameters of Eq. (\ref{eq:ads}) that lead  approximately to the properties of the  $X_{0}(3930)$ state, we obtain a shape for  $ \dfrac{d~\Gamma}{dM_{\rm inv} (D_{s}^{+} D_{s}^{-}) } $ compatible with the experiment. It is also welcome the fact that the strengths of the mass distributions at their peaks are of the same order of magnitude, same thing also found in the experiment. 

In summary, we have shown that a state below the$ D_{s} \bar{D}_{s} $ threshold, coupling
   strongly to $ D_{s} \bar{D}_{s} $ and more weakly to $D \bar D$, as found in the lattice QCD calculations, will necessarily produce a $ D_{s} \bar{D}_{s} $ mass distribution with a strong enhancement close to the $ D_{s} \bar{D}_{s} $  threshold. A quantitative evaluation of the $D^+ D^-$ and $D_s^+ D_s^-$ mass distributions in the $B^- \to D^+ D^- K^-$ and $B^- \to D_s^+ D_s^- K^-$ decays, with small modifications in the input used to obtain many hadronic states, shows that a $D_s^+ D_s^-$ bound state appears, which can be associated to the $X_0(3030)$, and this state, coupling both to $D \bar D$ and $D_s^+ D_s^-$, produces an enhancement in the $D_s^+ D_s^-$ mass distribution close to threshold with a shape in agreement with experiment. In addition, the relative strength of the $D^+ D^-$ and $D_s^+ D_s^-$ peaks is also similar to the experimental one. The conclusion is then that there is not need to invoke a new $X_0(3960)$ state, and the experimental observation is due to the presence of the $X_0(3930)$.

\section{ACKNOWLEDGEMENT}
This work is partly supported by the Spanish Ministerio
de Economia y Competitividad (MINECO) and European FEDER funds under Contract No. PID2020-112777GB-I00, and
by Generalitat Valenciana under contract PROMETEO/2020/023. This project has received funding from the European Union
Horizon 2020 research and innovation programme under the program H2020-INFRAIA-2018-1, grant agreement No. 824093
of the STRONG-2020 project. The work of A. F. was partially supported by the Generalitat 
Valenciana and European Social Fund APOSTD-2021-112, and the Czech 
Science Foundation, GA\v CR Grant No. 19-19640S.


\end{document}